\documentstyle[11pt,aaspp4,epsf,flushrt]{article}

\newcommand{\be}{\begin{equation}}
\newcommand{\ee}{\end{equation}}
\newcommand{\bgo}{\begin{eqnarray}}
\newcommand{\ego}{\end{eqnarray}}
\newcommand{\bg}{\begin{eqnarray}}
\newcommand{\eg}{\end{eqnarray}}
\newcommand{\ba}{\begin{eqnarray}}
\newcommand{\ea}{\end{eqnarray}}
\newcommand{\brr}{\begin{array}}
\newcommand{\err}{\end{array}}
\newcommand{\bc}{\begin{centre}}
\newcommand{\ec}{\end{centre}}

\newcommand{\h}{$h^{-1}$\thinspace}
\slugcomment{Accepted for publication in The Astrophysical Journal}
\lefthead{Pons--Border\'{\i}a et al.}
\righthead{Comparing estimators of the galaxy correlation function}
\begin{document}
\lefthead{Pons--Border\'{\i}a et al.}
\title{Comparing estimators of the galaxy correlation function}
\author{Mar\'{\i}a--Jes\'us Pons--Border\'{\i}a\altaffilmark{1}}
\affil{Departamento de F\'{\i}sica Te\'orica, Universidad Aut\'onoma de Madrid,
E--28049 Cantoblanco, Madrid, Spain}
\author{Vicent J. Mart\'{\i}nez\altaffilmark{2}}
\affil{Departament d'Astronomia i Astrof\'{\i}sica,
Universitat de Val\`encia, E--46100 Burjassot,
Val\`encia, Spain}
\author{Dietrich Stoyan and Helga Stoyan\altaffilmark{3}}
\affil{Institut f\"ur Stochastik, Freiberg University of Mining and Technology,
09596 Freiberg, Germany}
\and
\author{Enn Saar\altaffilmark{4}}
\affil{Tartu Observatory, T\~oravere, 61602, Estonia}

\altaffiltext{1}{Email:pons@astro1.ft.uam.es}
\altaffiltext{2}{Email: vicent.martinez@uv.es}
\altaffiltext{3}{Email: stoyan@orion.hrz.tu-freiberg.de}
\altaffiltext{4}{Email: saar@aai.ee}
\begin{abstract}
We present a systematic comparison of some usual estimators of the 2--point
correlation function, some of them currently used in Cosmology, others
extensively employed in the field of the statistical analysis of point
processes. 
At small scales, it is known that the correlation function
follows reasonably well a power--law expression $\xi(r) \propto r^{-\gamma}$.
The accurate determination of the exponent $\gamma$ (the order of the pole)
depends on the estimator used for $\xi(r)$; on the other hand, its behavior 
at large scale gives information on a possible trend to homogeneity.
We study the concept, the possible bias,
the dependence on random samples and the errors of each estimator.
Errors are computed by means of artificial catalogues of Cox processes
for which the analytical expression of the correlation function is
known.
We also introduce a new method for extracting simulated galaxy samples
from cosmological simulations.

\end{abstract}


\section{Introduction}

The two--point correlation function $\xi(r)$ has been
the primary tool for quantifying large--scale
cosmic structure (see Peebles 1980).
Several estimators have been used in the literature to
measure this statistical quantity from the redshift surveys.
The power--law shape of $\xi(r)$ seems to be well established
for $0.1 < r < 10 \, h^{-1}$ Mpc ($h$ being the Hubble
constant in units of 100 km s$^{-1}$ Mpc$^{-1}$):
\be
\xi(r) = \left ( r \over r_0 \right )^{-\gamma} .
\ee
However the reported values in the literature for the exponent $\gamma$
and the so--called correlation length $r_0$ (just related to
the amplitude $A$ of $\xi(r)=Ar^{-\gamma}$ by $A=r_0^{\gamma}$)
vary somewhat depending on the sample analyzed,
the estimator used, the weighting scheme, and the fitting procedure
employed.

Redshift-space distortions affect strongly
the correlation function at small scales;
the real-space correlation function
$\xi(r)$ is sometimes
derived from the $\xi(r_{\rm p},\pi)$ which depends on the radial
and projected separations.
For example, Davis \& Peebles (1983) found that for the CfA-I
redshift survey the values of the fit for the real-space correlation
function are consistent with $\gamma=1.77 \pm 0.04$ and
$r_0= 5.4 \pm 0.3$ \h Mpc. From the APM galaxy survey Maddox
et al. (1990) inferred that
$\gamma \simeq 1.66$ from measurements of the angular two--point
correlation function and the use of the Limber equation.
Other estimates of the two--point correlation
function in redshift space for the CfA (I and II) catalogues have produced
a variety of fits for $\xi(s) = (s/s_0)^{\gamma_s}$ (de Lapparent, Geller
\& Huchra 1988; Mart\'{\i}nez et al. 1993; Park et al. 1994)
with values for
$\gamma_s \sim 1.3 - 1.9$ and $s_0 \sim 4.5 - 12 \, h^{-1}$ Mpc.
For the Pisces--Perseus redshift survey Bonometto et al. (1994) found
$\gamma_s= 1.51 \pm 0.04$ and $s_0 = 7.4 \pm 0.7$ \h Mpc while, for
the SSRS, Maurogordato, Schaeffer \& da Costa (1992)
found $\gamma_s \simeq 1.6$ and $s_0 \sim 5 - 8.5$ \h Mpc.
Luminosity segregation and the presence of large scale inhomogeneities
affect the estimation of the parameters $\gamma_s$ and $s_0$ from the data
(Hamilton 1988; Davis et al. 1988; Mart\'{\i}nez et al. 1993).
In particular, for the
first slice of the CfA-II sample (de Lapparent et al. 1986) the two--point
correlation function shows a flatter shape with $\gamma_s \simeq 1.2$ and
$s_0 \simeq 10$ \h Mpc (de Lapparent et al. 1988; Mart\'{\i}nez et al. 1993).
Recent analyses of the shallower Stromlo--APM redshift survey performed
by Loveday et al. (1995) have provided fits for the redshift-space
correlation function ($\gamma_s \simeq 1.47$ and
$s_0 \simeq 5.9$ \h Mpc) and for the real-space
correlation function ($\gamma \simeq 1.71$ and
$r_0 \simeq 5.1$ \h Mpc). Regarding optical galaxies, it is worth
mentioning the best fitting values for $\xi(s)$ reported by Hermit et al.
(1996) for the ORS catalogue, $1.5 \le \gamma_s \le 1.7$ and $6.5 \le s_0 \le
8.8 $ \h Mpc, and the corresponding values for the derived
real space correlation function $1.5 \le \gamma \le 1.7$ and $4.9
\le r_0 \le 7.3 $ \h Mpc.

{\it IRAS} galaxies present typically a lower value of
the slope of the two--point correlation function: $\gamma \simeq 1.6$
(Davis et al. 1988; Saunders, Rowan--Robinson \& Lawrence 1992).
For the 1.2--Jy {\it IRAS} galaxy redshift survey, Fisher et al. (1994)
found that the parameters fitting the redshift space two--point correlation
function were $\gamma_s \simeq 1.28$
and $s_0 \simeq 4.53$  and for the derived
real space correlation function $\gamma \simeq 1.66$ and $r_0 \simeq
3.76$. These results are in agreement with the values obtained for the
QDOT-{\it IRAS} (1 in 6) redshift survey (Moore et al. 1994;
Mart\'{\i}nez \& Coles 1994).

It is however
important to have a good knowledge of the shape of the
two--point correlation function at small scales and in particular of the
value of $\gamma$ because it provides important
constraints on models of structure formation. The parameters obtained by
fitting the estimated two-point correlation function to a power--law
may depend on the estimator used to measure $\xi(r)$ from the redshift surveys.

In this paper we compare some of the estimators of $\xi(r)$ commonly
used in the literature concerning the large-scale structure of the Universe
and in the literature regarding the statistics of the
spatial point processes.
The paper is organized as
follows. We give the necessary
definitions in Section 2.
Section 3 illustrates the application of the estimators on galaxy samples
with different types of limitations.
In Section 4 we present a new method
for extracting artificial galaxies from simulations and
we introduce the so-called Cox processes. In Section 5
we perform the comparison of the given estimators under various
conditions (number of auxiliary random points used, number
of galaxies, etc). Our scheme to compute the errors of the
correlation function is introduced in Section 6 and
applied to the extracted synthetic galaxy samples.
Finally, in Section 7 we state our main conclusions.

\section{Estimators of the correlation function}

In the framework of the statistical analysis of the large scale
structure of the Universe, one assumes that the three--dimensional
point pattern of
galaxies is a sample of a stationary and isotropic point
field. For such a point field the intensity $\lambda$ is the first order
characteristic; $\lambda$ equals the mean number of points per unit volume.
Second order characteristics are the correlation function $\xi(r)$ and the
pair correlation function $g(r)$, which satisfy
\bgo
g(r)=1+\xi(r).
\ego
The function $g(r)$ is defined as follows. Consider an infinitesimal
ball $B$ of volume d$V$. The probability of having a point of the point field
in $B$ is $\lambda{\rm d}V$. If there are two such balls $B_1$ and $B_2$ of
volumes d$V_1$ and d$V_2$ and inter-centre distance $r$ then the
probability to have a point in each ball can be denoted by $P(r)$. It
can be expressed as
\bgo
P(r)=g(r)\lambda{\rm d}V_1\lambda{\rm d}V_2\,.
\ego
The factor of proportionality $g(r)$ is the pair correlation function. It is
clear that, in the case of complete randomness of the point distribution,
$g(r)=1$.

For statistical estimation of $\xi(r)$, $N$ points are given inside
a window $W$ of observation,
which is a three--dimensional body of volume $V$.

Several estimators of $\xi$ are commonly used. The most
extensively used one
is that of Davis \& Peebles (1983), for which an auxiliary random
sample containing $N_{\rm{rd}}$ points must be generated in $W$ and the
following quantity must be computed:
\begin{equation}
\hat\xi_{\rm{DP}} (r) = \frac{DD(r)}{DR(r)}\times \frac{N_{\rm{rd}}}{N} -1,
\end{equation}
where $DD(r)$ is the number of all pairs 
in the catalogue (window $W$) with
separation \lq\lq close to $r$'', i.e., inside the interval
$[r-dr/2, r+dr/2]$, and $DR(r)$ is the number of
pairs between the data and the random sample with separation
in the same interval. The symbol $\hat{}$ on top of an statistical 
quantity denotes  its estimator.
For flux--limited samples one has to weight
each galaxy by means of the inverse of the selection function; since
we basically deal in this paper with complete samples, this will not be
considered here.

Another possibility is to use the estimator proposed by Hamilton (1993), which
has become very popular since its introduction and reads:
\begin{equation}
\hat\xi_{\rm{HAM}} (r) = {\frac{DD(r)\times RR(r)} {[DR(r)]^2}} - 1
\label{eham},
\end{equation}
where also the number of pairs in the random catalogue with separation
in the interval mentioned above, $RR(r)$, is taken into account.
Hamilton (1993) has shown that the dependence of $\hat\xi_{\rm{HAM}}$
on the uncertainty in the
mean density is of second order, while in $\hat\xi_{\rm{DP}}$
it is linear and presumably dominates at large scales.
He also considers the accurate computation of $RR$ and $DR$ by a combination
of analytical and numerical integration, decomposing
the separations into their radial and spatial parts.

One more estimator was proposed simultaneously (in the literal
sense of the word\footnote{Both papers were received in ApJ the very same
day.}) to Hamilton's, by Landy\& Szalay (1993):
\begin{equation}
\hat\xi_{\rm{LS}} (r) = 1 + \frac{DD(r)}{RR(r)}\times\left(\frac{N_{\rm{rd}}}
{N}\right)^2 - 2 \frac{DR(r)}{RR(r)}\times \frac{N_{\rm{rd}}}{N}\label{els}.
\end{equation}
Szapudi \& Szalay (1997) claim that LS behaves like HAM except for a
small bias.

A different kind of estimator was introduced by Rivolo (1986), in which
random samples do not explicitly appear:
\begin{equation}
\hat\xi_{\rm{RIV}} (r) = \frac{V}{N^2} \sum_{i=1}^N \frac{n_i(r)}{V_i} -1,
\end{equation}
where $n_i(r)$ is the number of neighbours
at distance in the interval $[r-dr/2, r+dr/2]$ from galaxy $i$ and
$V_i$ is the volume of the intersection
with $W$ of the shell centred at the $i$th galaxy and having
radii $r-dr/2$ and $r+dr/2$.
In the case of $W$ being a cube, an analytic
expression for $V_i$ is provided in Baddeley et al. (1993).
By the way, $\hat\xi_{\rm {RIV}}$ is closely related to
Ripley's estimator of the so-called $K$--function, which is an integral of
the correlation function $g(r)$ (Ripley 1981; Stoyan \& Stoyan 1994;ç
Kerscher 1998).

Before introducing a fifth estimator $\hat\xi_{\rm{STO}}(r)$,
which is commonly used in the
framework
of spatial point processes,
let us define
a naive estimator $\varrho^*(r)$ of the product density
$\varrho(r)=\lambda^2g(r)$:

\bgo
\varrho^*(r)=\frac{DD(r)}{4\pi r^2 dr V}\,.
\ego
The estimator of $\xi(r)$ is then
\bgo
\xi^*(r)=\frac{\varrho^*(r)}{\hat\lambda^2}-1=\frac{DD(r)/N}{4\pi r^2
  dr \hat\lambda}\, ,
\ego
with $\hat\lambda={} N/V$.

A smoothened version
$\widetilde{DD}(r)$ of $DD(r)$ can be obtained
by means of a
kernel function $k(x)$. Here the Epanechnikov kernel is used
\bgo
k(x)=\left\{\begin{array}{ll}\frac{3}{4w}\big(1-\frac{x^2}{w^2}\big)
\quad&\mbox{for }|x|\le w\\
0&\mbox{otherwise}\end{array}\right.\,.
\ego
The parameter $w$ is called bandwidth.
Now $\widetilde{DD}(r)$ is
\bgo
\widetilde{DD}(r)=\sum\limits_{i=1}^N\sum\limits_{\left.j=1\atop
i\ne j\right.}^N k(r-|{\mathbf x}_i-{\mathbf x}_j|)\,,
\ego
where ${\mathbf x}_i$ is the location of the
$i$th galaxy in $\bf{R}^3$
and those pairs with distances close to $r$ will
contribute to the sum. Of course, the vagueness of the expression ``close
to $r$'' is not completely overcome by means of the kernel function;
the choice of the bandwidth $w$ is an art (see below).

\begin{figure}
\begin{center}
\epsfxsize=6.cm
\begin{minipage}{\epsfxsize}\epsffile{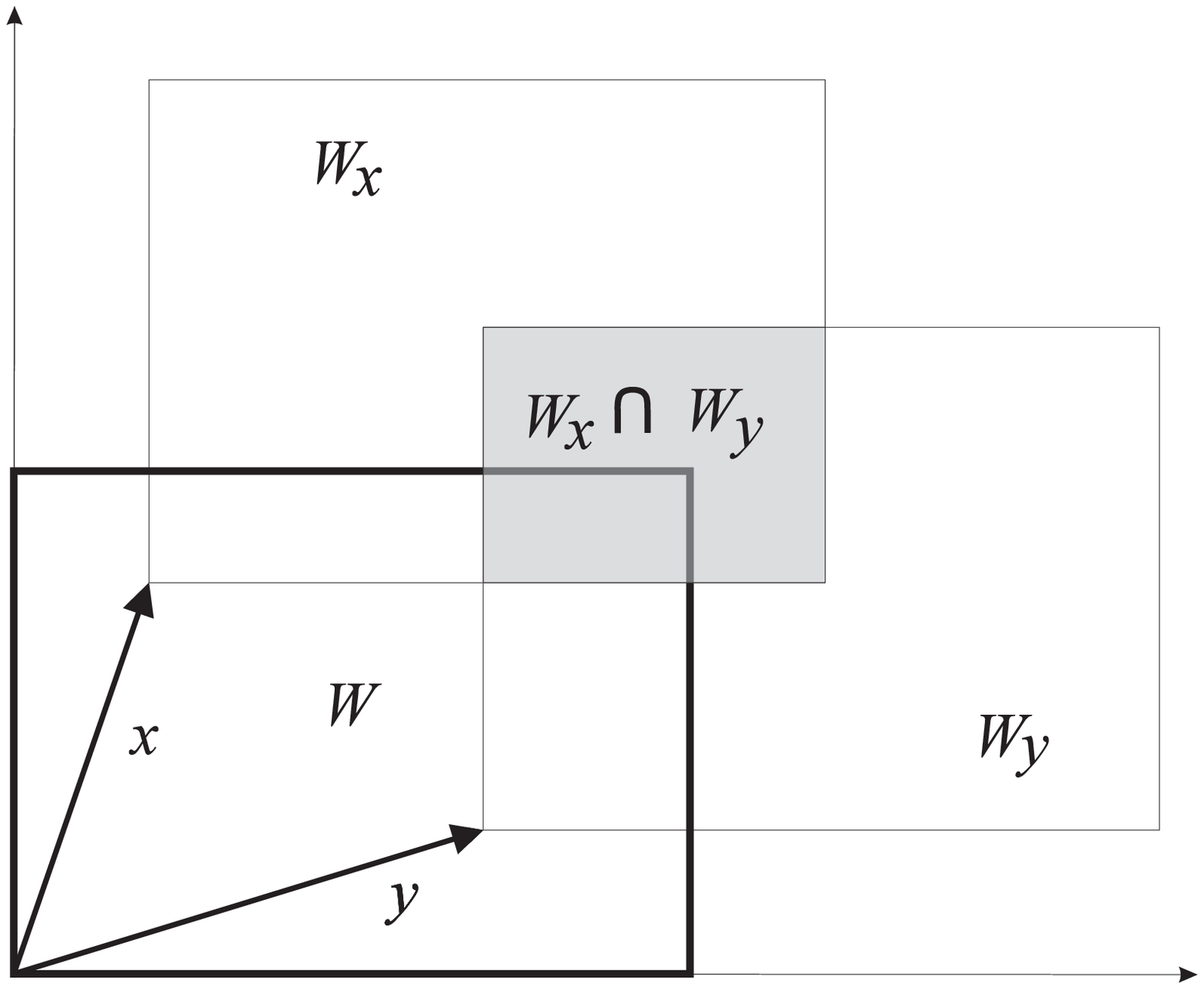}\end{minipage}
\caption{A 2D representation of the denominator in Eq. 12.}
\label{win}
\end{center}
\end{figure}

A serious drawback of the naive estimator $\varrho^*(r)$ is that it is not
edge--corrected and certainly there are edge--effects: points close to the
boundary of $W$ do not find as many neighbours as points in the inner region
of $W$ do. Thus $DD(r)$ or $\widetilde{DD}(r)$ tends to be smaller than
expected and the estimator $\varrho^*(r)$ produces too small values.
Let us remark that the
problems with edge--effects in three--dimensional space are
much more serious
than in one-- and two--dimensional space, which is typical
in many fields of spatial statistics:
for a square of unit side length the fraction
of the area wasted by a buffer zone of width 0.1 would be 36~\%, while
the fraction of the volume in a unit cube would be 48.8~\%. Consequently,
careful edge--correction is necessary. Various forms of
doing it are presented in Stoyan
and Stoyan (1994) for planar point processes. Here
a form is used which is suitable for the case of homogeneous (not
necessarily isotropic) point fields and it yields an unbiased estimator of
$\varrho$, which reads:
\bgo
\hat\varrho_{\rm{STO}}(r)=\frac{1}{4\pi r^2}\sum\limits_{i=1}^n\sum\limits_
{j=1\atop j\ne i}^n\frac{k(r-|{\mathbf x}_i-{\mathbf x}_j|)}
{V(W\cap W_{{\mathbf x}_i-{\mathbf x}_j})}\label{estim},
\ego
from which we have that
$\hat g_{\rm{STO}}(r)=1+\hat\xi_{\rm{STO}}(r)=
\hat\varrho_{\rm{STO}}(r)/\hat\lambda^2$.
Here $W_{\mathbf y}$ denotes the window $W$ shifted by the vector
${\mathbf y}$, $W_{\mathbf y}=W+{\mathbf y}=\{{\mathbf x}:{\mathbf x}={\mathbf
z}+{\mathbf y},{\mathbf z}\in W\}$.
The denominator is the volume of the window intersected with a version
of the window which has been shifted by
the vector ${\mathbf x}_i-{\mathbf x}_j$ and
it can be written
also as $W_{{\mathbf x}_i}\cap W_{{\mathbf x}_j}$ (see Fig.\ref{win}).
Clearly, this
volume is smaller than the window volume which
appears in the naive estimator; thus edge--correction is done.

We want to emphasize here that the point process does not need to be
isotropic to get good estimates of $\xi(r)$ through
$\hat\xi_{\rm {STO}}$, contrary to the four previously
mentioned estimators. This property of
the $\hat\xi_{\rm {STO}}$ estimator makes it very useful, especially when
measuring  the correlation function in redshift space, because
peculiar motions act to erase small scale correlations,
flattening thus the shape of the correlation
function and providing smaller values for $\gamma$.
Beware applying statistics which are not suitable for anisotropic
processes, since experience shows that
deviations from isotropy may cause great errors if isotropic case estimators
are used. In such cases, one can improve the STO estimator by replacing
$4 \pi r^2$ in the denominator of Eq.~\ref{estim} by the quantity
$4 \pi |{\mathbf x}_i-{\mathbf x}_j|^2$.

The estimator $\hat\xi_{\rm {STO}}$ uses a smoothing kernel in order
to reduce shot noise.
The problem of shot noise arises, especially, with DP and HAM estimators
because, 
at small scales,
$DR(r)$ becomes very
small due to the fact that the number of Poisson points within a shell of
radius $r$ is approximately  proportional to $r^2$.
It is worth mentioning that
Davis \& Peebles (1983) already tried to reduce the shot noise
by smoothing $DR(r)$ at small scales ($r < 2$ \h Mpc).
Other authors change the estimator
used at small scales (van de Weygaert 1991). Other solutions to this problem
will be commented in section 5.

There is still another well-known but little appreciated
(Blanchard \& Alimi 1988) estimator introduced for the study of the
angular correlation function by Peebles \& Hauser (1974). Its
three--dimensional
counterpart is

\bgo
\hat\xi_{\rm {PH}} =
   \frac{DD(r)}{RR(r)} \times
   \left(\frac{N_{\rm {rd}}}{N}\right)^2 - 1 \label{eph}.
\ego

Peacock (1992) argues that $\hat\xi_{\rm {PH}}$ and $\hat\xi_{\rm {DP}}$
should be equivalent when applied to a large volume; however the latter is less
sensitive to whether there is a rich cluster close to the border of the sample.

It can be shown (Kerscher 1998) that $\hat\xi_{\rm {PH}}$ is nothing else than
the isotropized Monte Carlo
counterpart of  $\hat\xi_{\rm {STO}}$ in
which the smoothing kernel has been substituted by the standard count of pairs
$DD(r)$.

The relation between the estimators LS and PH can be easily deduced
from their definitions given in Eqs.~\ref{els} and \ref{eph},

\bgo
\hat\xi_{\rm {LS}} = \hat\xi_{\rm {PH}} + 2 -2
\left ({DR(r) \over RR(r)}\times{N_{\rm rd} \over N} \right )\label{phls}.
\ego

In a broad sense, most of the estimators consist of a sum of pairs
in the numerator whereas the denominator is an edge-corrected version of the
denominator in $\varrho^*(r)$. The differences among them
lie essentially in the way of performing this border
correction in the denominator. In cosmology we have to cope
most often with complicated windows, so the calculation of $RR$ and $DR$
has to be performed through Monte Carlo integration.

Within this general scheme,
RIV presents (at first sight) a certain deviation by summing means of
edge-corrected counts of pairs instead of summing the
means first and dividing them afterwards like the other
estimators do. 
HAM and STO both 
present a new approach to
the problem: the former arises from minimizing the dependence of the
variance on the (not always well known) intensity and the latter
introduces a smoothing in the counting of pairs of galaxies.

The estimator $\hat\varrho_{\rm {STO}}(r)$ has an irregularity property
for small $r$
resulting from the denominator $4\pi r^2$. If the numerator of
$\hat\varrho(r)$ vanishes, then $\hat\varrho_{\rm {STO}}(r)=0$
by definition. But if
there is at least one pair with a very small interpoint distance then the
numerator is positive and $\hat\varrho_{\rm {STO}}(r)$
may take a very large value.
This problem is discussed in Stoyan and Stoyan (1996). For many point fields
this effect does not play a role and it suffices to avoid too small values of
$r$. However, in the case of galaxies $\xi(r)$ is known to have a pole at
$r=0$ and the multiplicity of this pole is the value of the exponent $\gamma$.
Thus small values of $r$ are important and it is precisely in this
region where we can observe remarkable differences among the various estimators
considered. The set of small $r$ values is not an easy zone to study clustering
in because at small distances there are few
points and the shot noise dominates; consequently it would be interesting to
check if any of the estimators is able to cope at least moderately well with
this kind of noise.

On the other hand, in the STO case a contrary effect
influences the estimation problem,
namely the fact that
kernel estimators tend to smooth the results. This
may lead to values of $\hat\varrho(r)$ which are too low for small $r$.
Stoyan and Stoyan (1996) recommend to use large samples and small values of the
bandwidth $w$, taking numerical experiments with statistical data from
simulated point fields in order to find out the best value; such
experiments have led to the result that a good choice
of $w$ would be
\bgo
w\equiv c\lambda^{-1/3}\,
\ego
with the coefficient $c$
being around 0.1 for point fields such as the Poisson
point process. For cluster processes, values of $c$
around 0.05 
have yielded acceptable
estimates of $\xi(r)$ also for small $r$ and this is the value we use
throughout the paper.

\section{The estimators acting on galaxy samples}

The aim of this Section is to stress the fact
that there exists no ``perfect
estimator'' but that, as Doguwa \& Upton (1986) remark,
the usefulness of an estimator can depend on the
kind of process/sample/distance range under study.

\subsection{Comparison between DP and HAM}
The currently most widely used estimators in the literature are
DP and HAM. In this Section we are going to perform a comparison
between them by analyzing results of applying them to galaxy samples
which have been obtained in different ways.

\subsubsection{Complete volume--limited samples}
\begin{figure}
\begin{center}
\epsfbox{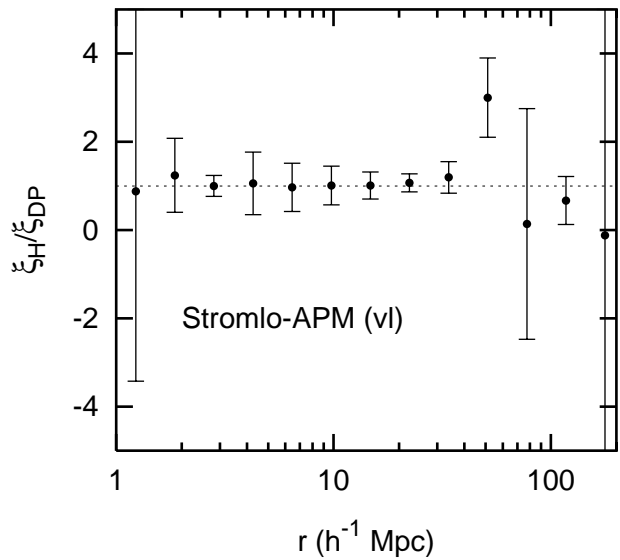}
\caption{The quotient between the Hamilton and Davis \& Peebles
estimators for the correlation function of a volume--limited
sample extracted from the Stromlo-APM  redshift survey.}
\label{difapm}
\end{center}
\end{figure}

We plot in Fig.~\ref{difapm} the quotient between the Hamilton and the Davis
\& Peebles estimators of the correlation function for a
volume--limited sample extracted from Stromlo-APM, where the values of the
correlation functions and of bootstrap errors have been
provided to us by J.~Loveday. In that case the relative differences
are again small and much less significant than the bootstrap errors.
In fact this result is used by Loveday et al. (1995) to clarify
a possible concern regarding the HAM estimator, showing that it does not
remove intrinsic large-scale clustering.
So it seems that the main difference between both estimators happens
when they are applied to a sample whose density is poorly
known, where HAM works better.
This is a very sparse sample (only 1 in 20 galaxies from the angular
sample is included in the redshift survey), therefore at small scales
the statistical quantities are rather noisy.
It is interesting to note that the value of $\xi(r)$ at $r=1.23 \, h^{-1}$
Mpc is 2.7 for DP and 2.4 for HAM. Although at this scale the error bar
is quite large (between 5 and 10), it is clear that the value of $\xi(r)$, 
assuming it follows a power law,
is underestimated by both estimators, indicating a strong bias. At the
same scale the RIV estimator provides a larger value for $\xi(r)$, 13.6,
which clearly is more acceptable.

\subsubsection{Samples with non-uniform density}

The expressions we have presented for the estimators
are adequate for samples which are either complete or
volume--limited. They can be generalized to other kinds
of limitation by assigning to each galaxy weights
inversely proportional to a certain selection function.
This function represents the fraction of the total
population of galaxies satisfying the limitation
criterion at a certain distance. The weighting scheme
used or the uncertainty in the knowledge of the
selection function can influence, however, the result
for the correlation function. Since we want to compare
estimators this added uncertainty would disturb
unnecessarily the measure, so we shall mainly work with
complete or volume--limited samples.

\begin{figure}
\begin{center}
\epsfbox{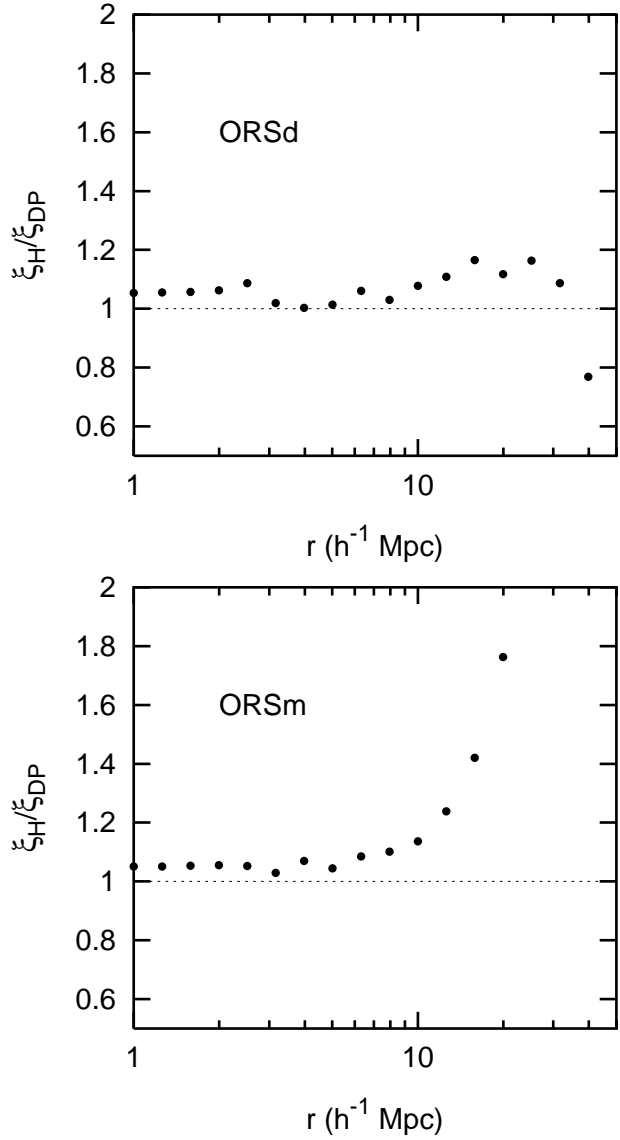}
\caption{The quotient between the Hamilton and Davis \& Peebles
estimators for the correlation function of samples extracted from
the Optical Redshift Survey (top: diameter--limited and
bottom: magnitude--limited).}
\label{difors}
\end{center}
\end{figure}

Nonetheless, we want to show
briefly in this subsection an example of the difference of
applying DP and HAM to incomplete samples. In particular we have
used two samples extracted from the Optical
Redshift Survey (described in Santiago et al. 1995), one limited in apparent
magnitude and the other in diameter.
What we show in Fig.~\ref{difors} is the quotient between
both estimators, i.e., $\hat\xi_{\rm{HAM}} /\hat\xi_{\rm{DP}}$,
calculated by Hermit et al. (1996).
The differences are only noticeable at very large scales and they are
bigger for the magnitude--limited sample (bottom panel) than for the
diameter--limited sample (upper panel).
This fact is remarkable because the
latter sample is sparser at large distances
than the former,
since the selection function is
steeper for the diameter--limited sample than for the
magnitude--limited one (Santiago et al. 1996). However, the Galactic
extinction affects galaxy magnitudes more strongly than diameters.
The selection function used by Hermit et al. (1986) incorporates
an angular dependence modelling the extinction and this fact could explain
the deviations observed in Fig.~\ref{difors}.
In fact for the Las Campanas redshift survey, having a very complex 
selection function,
Tucker et al. (1997) have shown that, at large scales, the differences
between both estimators can be as larger as the signal itself.

\subsection{The six estimators acting on a volume--limited sample}

\begin{figure}
\begin{center}
\epsfbox{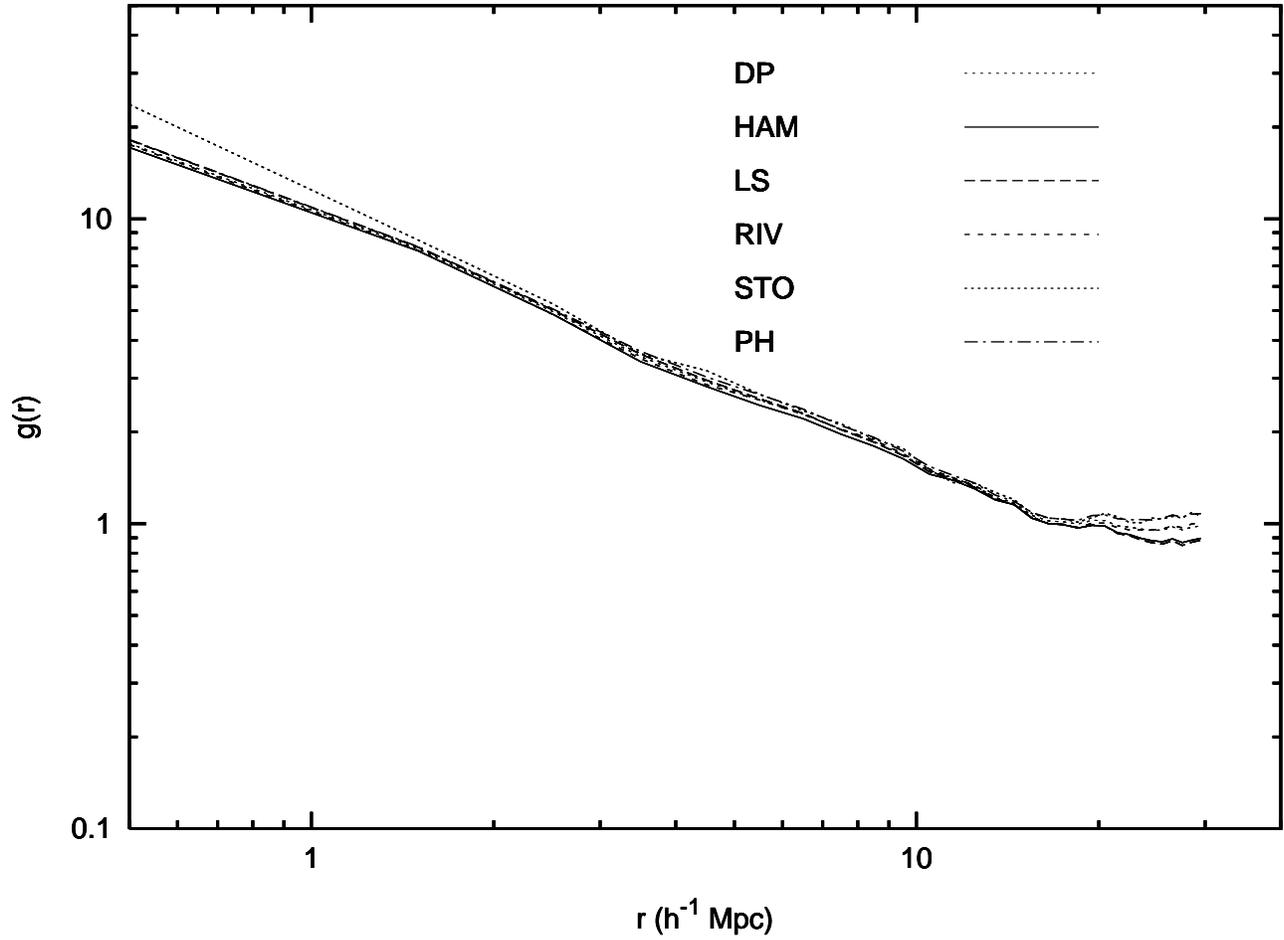}
\caption{The correlation function of the PPS sample calculated
by means of the six estimators described in Section 2.}
\label{pps}
\end{center}
\end{figure}

Now we shall apply the six mentioned estimators to a complete sample,
volume--limited to 79\h Mpc, extracted from the Perseus-Pisces
Survey (for a thorough description of the sample, see Kerscher et al.
1997). The results can be observed in Fig.~\ref{pps} and show that, at
small and intermediate scale, all estimators behave similarly
except STO, which gives a bigger value of $g$; as
we shall later see, this estimator has a smaller variance than the
others at small scales, important for the determination of $\gamma$. 
This can be interpreted saying that its nature makes it less sensitive 
to local anisotropies due to peculiar motions.
This result mainly indicates
that all the estimators measure the two-point correlation function
rather well in the ``easy'' range $2 < r < 15$\h Mpc. For bigger scales, 
relevant for information on a possible trend to homogeneity of the matter 
distribution, there are some differences as well.
Therefore, it is worth to study
the behaviour of the different estimators on controllable point sets
in order to know the deviation of each one from the true value of the
two-point correlation function and the ensemble
variance. The test performed in Section 5 points in
this direction.

\section{Description of the artificial samples}

\subsection{Cox processes}

\begin{figure}
\hspace{.6cm}
\epsfbox{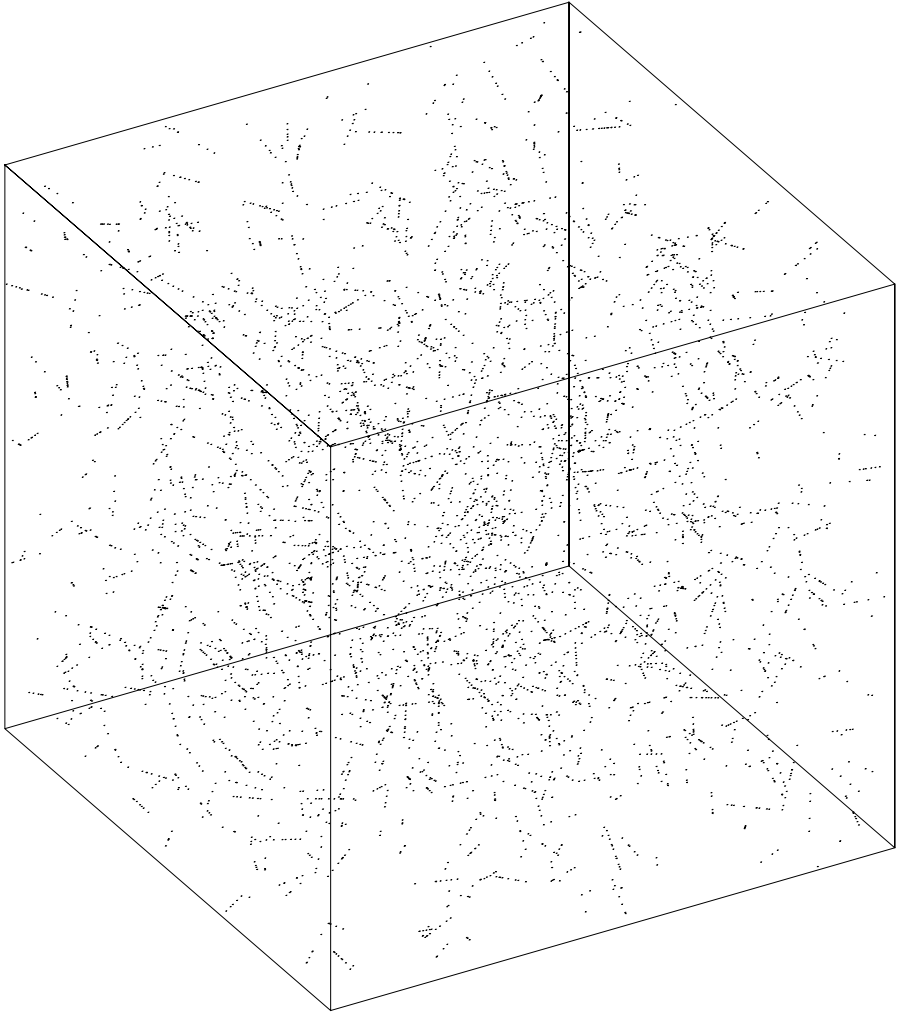}
\caption{Simulation of the Cox process with 
$N=6007$ points.}
\label{coxsim}
\end{figure}

We shall make use of an artificial
sample which is a particular kind of a so-called segment Cox point process.
This is a clustering process for which an analytical expression of its
2--point correlation function is known and therefore can be used as a
test to check the accuracy of the $\xi$--estimators.
The variant we are going to use is produced in the following way: segments of
length $l$ are randomly scattered inside a cube $W$ (see Fig.~\ref{coxsim}) and
on these segments points are randomly distributed.
Let $L_V$ be the length
density of the system of segments, $L_V=\lambda_{\rm{s}}l$, where
$\lambda_{\rm{s}}$ is
the mean number of segments per unit volume. If $\lambda_l$ is the mean
number of points on a segment per unit length, then the intensity $\lambda$
of the resulting point process is
\bgo
\lambda=\lambda_lL_V=\lambda_l\lambda_{\rm{s}}l\,.
\ego

For this point field the correlation function can be easily calculated
taking into account that the point field has a
driving random measure equal to the random length measure of the system of
segments. Stoyan, Kendall and Mecke (1995) have shown that the
pair correlation function of the point field equals the pair correlation
function of the system of segments, which reads
\bgo
\xi_{\rm {Cox}}(r)=\frac{1}{2\pi r^2L_V}-\frac{1}{2\pi rlL_V} \label{skm}
\ego
for $r \le l$ and vanishes for larger $r$. As we can see, the expression
is independent of the intensity $\lambda_l$.

In Section 5.2 we shall use 10 realizations of a segment Cox process
generated inside a cube of sidelength $L=100$ $h^{-1}$ Mpc 
with values of the
parameters $\lambda_{\rm{s}}=10^{-3}, \lambda_{\rm{l}}=0.6$, 
and $l=10$ $h^{-1}$ Mpc, which 
produces sets containing $N\simeq 6000 $ points.

\subsection{Simulated galaxies}

\begin{figure}
\hspace{1.cm}
\epsfbox{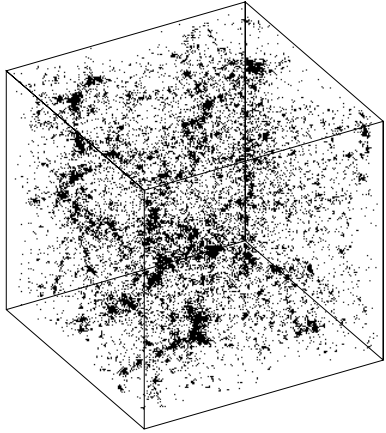}
\caption{Density distribution for the P$^3$M model.}
\label{Den-p3m}
\end{figure}

In this subsection we show how
a sample of synthetic galaxies was obtained
from a simulation of a CDM--type Universe.
The cubic region modeled was of sidelength 80\h Mpc,
a standard $\Omega=1$ Universe was chosen, and the
initial computational grid was 32$^3$, with the same number
of particles. The run started from small perturbation
amplitudes and was terminated when the
$\sigma_8$ parameter, the
mass dispersion in 8\h Mpc radius spheres,
was close to the observed value 1.
We used H.~Couchman's public
domain adaptive P$^3$M code (which can be obtained at
http://coho.astro.uwo.ca/pub/ap3m/ap3m.html), and the initial
data were those of the test model supplied with this
code. The initial
density perturbation spectrum was close to the
observed one for scales of 8--10\h Mpc with
a rather sharp cutoff used to eliminate numerical
effects:
\be
P(k)\sim k^{-1}\exp(-(k/k_c)^{16}) \label{power}.
\ee
The cutoff wavenumber $k_c=0.96h$Mpc$^{-1}$
is lower than the
Nyquist frequency used in the computations (with
a 32$^3$ grid the smallest usable wavelength is
5\h Mpc, while the cutoff wavelength is 6.5\h Mpc).
The final state of the model represents a continuous
distribution of dark matter in the computational
volume (see Fig.~\ref{Den-p3m}).

In order to get closer to observations one has to predict
the positions of
luminous objects (galaxies, their groups or clusters)
on the basis of this distribution. There
exist many essentially phenomenological methods for doing this,
and we have applied another one, the recent equal--mass binary tree approach.
These trees are known as multidimensional $k$-trees;
they were used first in the statistics of cosmological data
by van de Weygaert (1988) and have now been resurrected by
Suisalu et al.  (1999),
who give in that paper the
detailed description of their motivation and of the intricacies
of their use. The present application is ideal for these trees, having a
perfectly shaped volume and a number of particles that is a power of 2.

\begin{figure}
\epsfbox{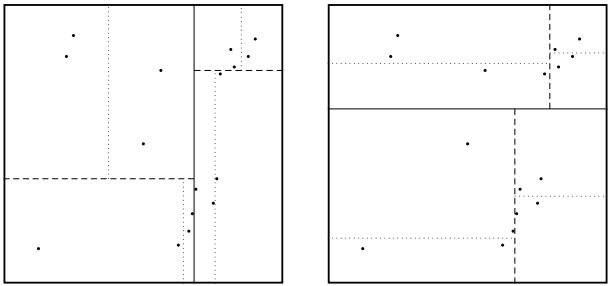}
\caption{Dividing up a surface keeping equal mass in each cell at a given level.
Different levels are represented with different line styles. Each panel shows
the division with different starting direction.}
\label{tree}
\end{figure}

The equal--mass trees are constructed by
dividing the sample volume successively into smaller
subvolumes, keeping the mass (number of points) of the two
subvolumes equal. In order to illustrate the method, we show in
Fig.~\ref{tree} how a planar point
process with $2^4$ points is divided by means of the equal--mass
tree for the two different starting directions.

\begin{figure}
\epsfbox{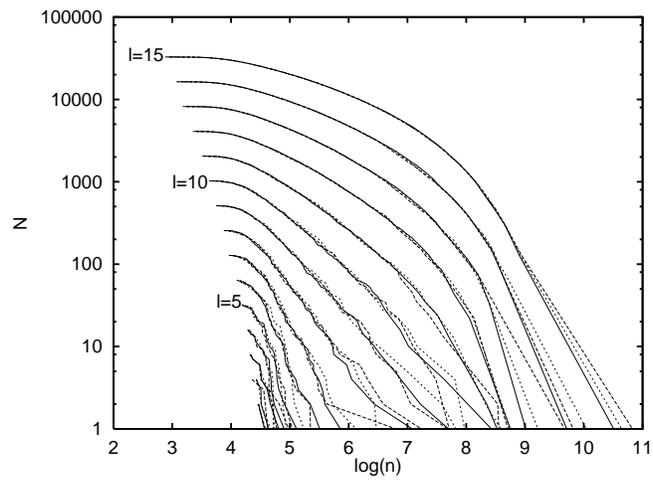}
\caption{Number of objects for a given level and density threshold.}
\label{Num-den}
\end{figure}

This procedure assigns a fixed mass to
a given level of subdivision, while the values of the
subvolumes and their positions describe the density distribution
for a given mass scale. One can select objects applying either a
mass or a density bias and we choose the latter.
In other words, for a given level of subdivision
all cells have the same mass, but different density. The density
is just proportional to the inverse of the volume of the cell.
The mass within a cell will form a galaxy if its density exceeds
a given threshold. We have applied this procedure to the CDM
simulation. In Fig.~\ref{Num-den} we have plotted the number of
cells $N$ with density exceeding a given density threshold $n$
for each level $l$ of subdivision.
It can be seen that the isolevel lines split into
three, showing the scatter for trees that have different starting
directions.

For the present study we used samples selected on the basis
of a fixed threshold density, $n=10^6$ (in units of number of
points divided by the fraction of the whole volume occupied by
the cell), and for four levels. Each
level can be assigned a fixed mass, $M_{\ell}=1.4\times
10^{17}h^{-1}2^{-\ell}M_\odot$.
The mass range for our samples runs from
$4.3\times10^{12}h^{-1}M_\odot$ for the finest subdivision,
somewhat higher than the total mass of a giant galaxy, to
$3.4\times10^{13}h^{-1}M_\odot$, characteristic for a group
or a poor cluster of galaxies. Each object gets its coordinates
from the centre of the cell that collapsed to form it, and we used
a fixed starting direction to construct a tree.

The spatial distribution of the objects of our samples is
shown on the left side of Fig.~\ref{Sim-gal}. From top to bottom the panels
correspond to levels $\ell=12,13,14,15$ and the number of points of
each subsample is respectively $N=762,1930,4734,11284$.
As it can be seen, the geometry of the mass
distribution for different mass levels does not differ much.

\section{Comparison of the estimators of the correlation function}

\subsection{Dependence on $N$}

\begin{figure}
\hspace{1.3cm}
\epsfbox{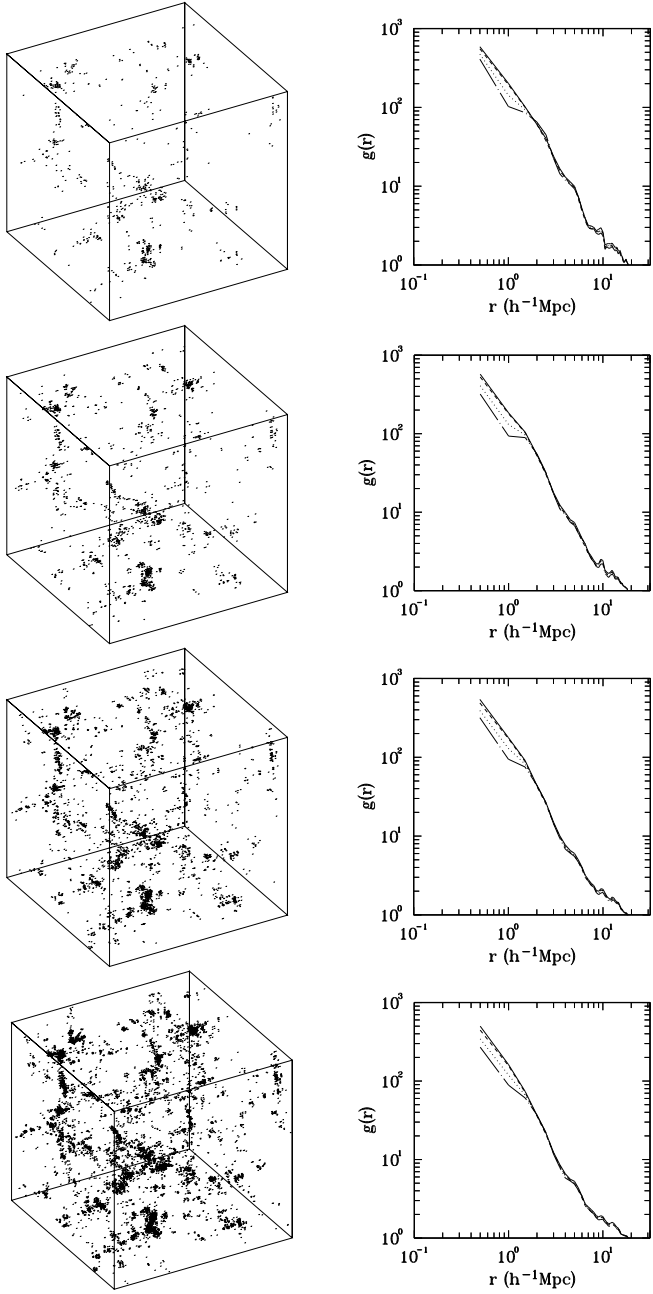}
\caption{Letf panels: spatial distributions of galaxies drawn from a
CDM simulation by means of the multidimensional binary tree. Right panels:
The corresponding function $g(r)$ for the four samples
calculated by means of four of the methods:
dotted line (Davis \&  Peebles), dashed-dotted line (Hamilton),
dashed line (Rivolo) and solid line (Stoyan \& Stoyan)}
\label{Sim-gal}
\end{figure}

We have calculated the pair correlation function $g(r)=1+\xi(r)$ for
the four samples shown in Fig.~\ref{Sim-gal} by means of four of the estimators
described in Section 2. Our aim was to check the influence of the total number
of points $N$ on each
of them. The extracted galaxies we have described in Section 4.2 are
appropriate
for this check because these samples trace the {\sl same} structure
with increasing number of points for bigger levels $\ell$.

The results are shown in the right panels of
Fig.~\ref{Sim-gal}. We can see that at large scales there is full
agreement among the four methods but, at short distances,
STO and RIV still agree rather well, while DP and HAM
deviate from this behaviour.
In all cases we have used random realizations
containing $N_{\rm{rd}} = 20000$ points each.
This is a typical number of random points used in the
computation of $\xi(r)$ (Dalton et al. 1994, Tucker et al. 1997).
We see in the plot that the relation among the
different estimators remains similar from one panel to
the other although $N$ is varying by a
factor 15 in total.

The conclusion is that $N$, provided it is big enough
to trace satisfactorily the main structures present in
the sample,  does not have a significant
influence on the estimation of the correlation function.

We have repeated this analysis
by using the same data sample ($\ell$ =12, $N= 762$ simulated galaxies), 
but different realizations of random samples (different seeds). 
For $10^4$ random points the differences in
correlation functions were appreciable for all four estimators that
use auxiliar random samples,
but for $10^5$ points the correlation functions practically coincided,
except for small $r$ values for DP and HAM.
In next subsection we study in more detail the dependence on $N_{\rm {rd}}$
by means of the Cox processes.

\subsection{Dependence on $N_{\rm {rd}}$}

\begin{figure*}
\begin{center}
\epsfxsize=6.5cm
\begin{minipage}{\epsfxsize}\epsffile{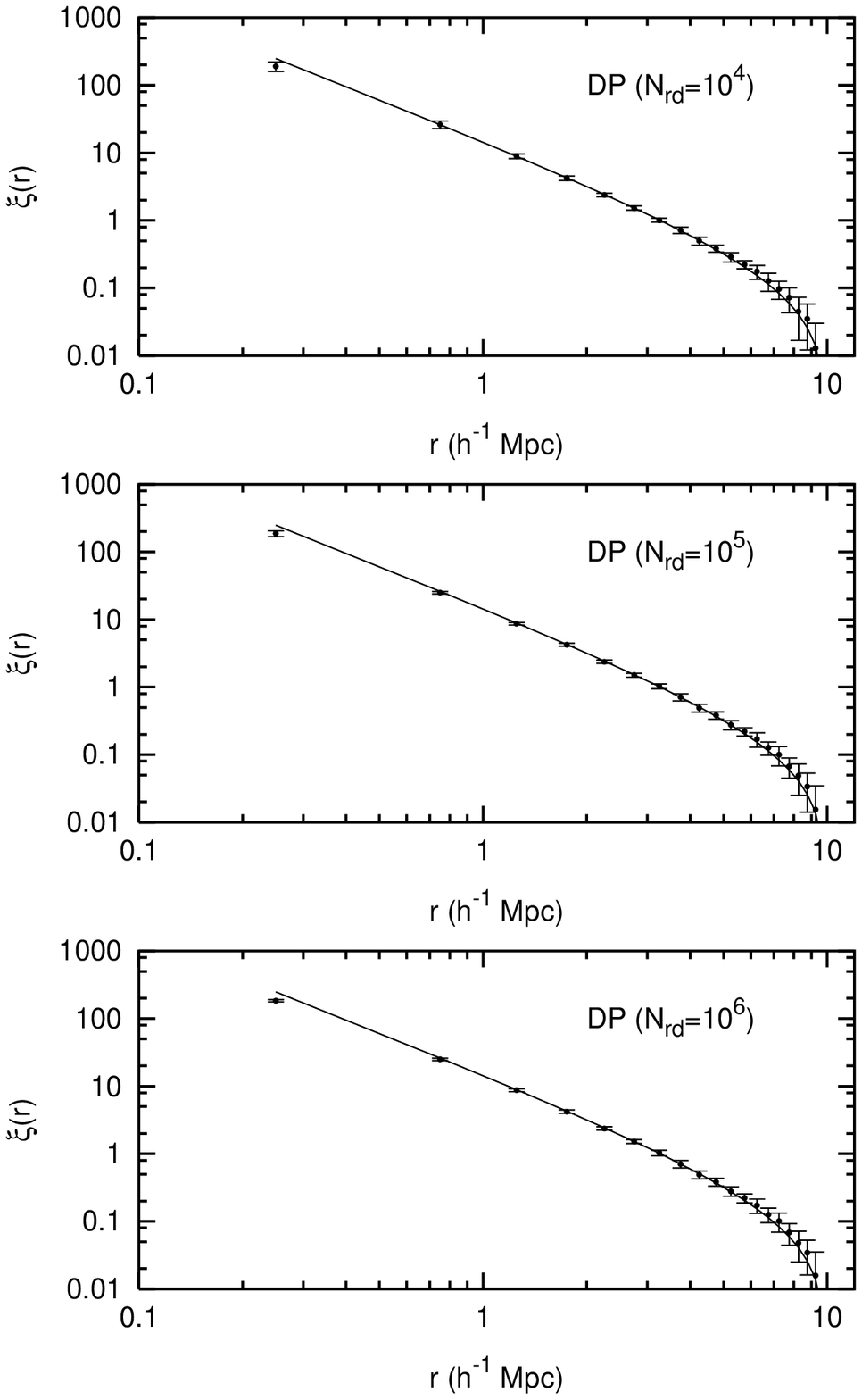}\end{minipage}
\epsfxsize=6.5cm
\begin{minipage}{\epsfxsize}\epsffile{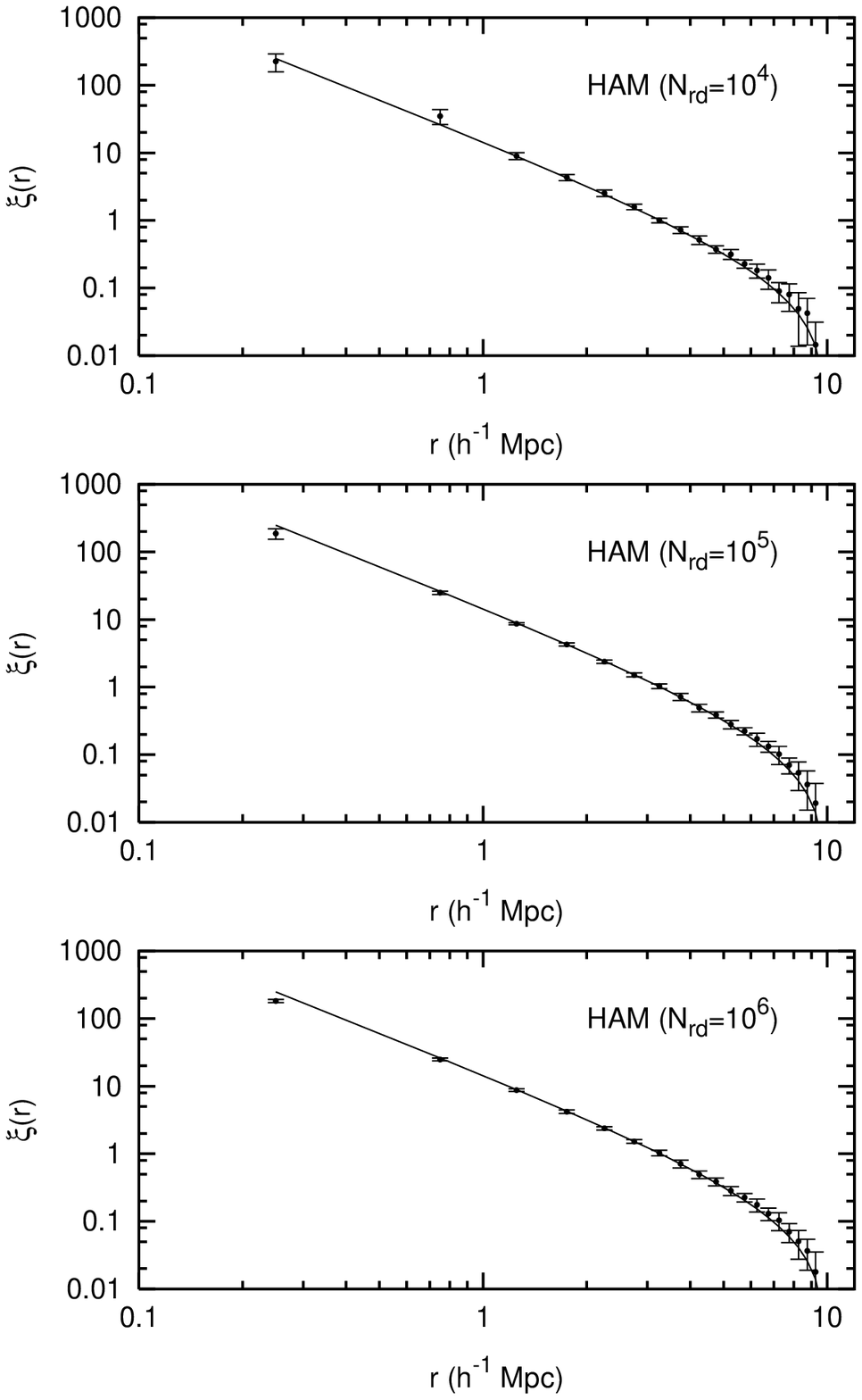}\end{minipage}
\epsfxsize=6.5cm

\vspace{1.cm}

\begin{minipage}{\epsfxsize}\epsffile{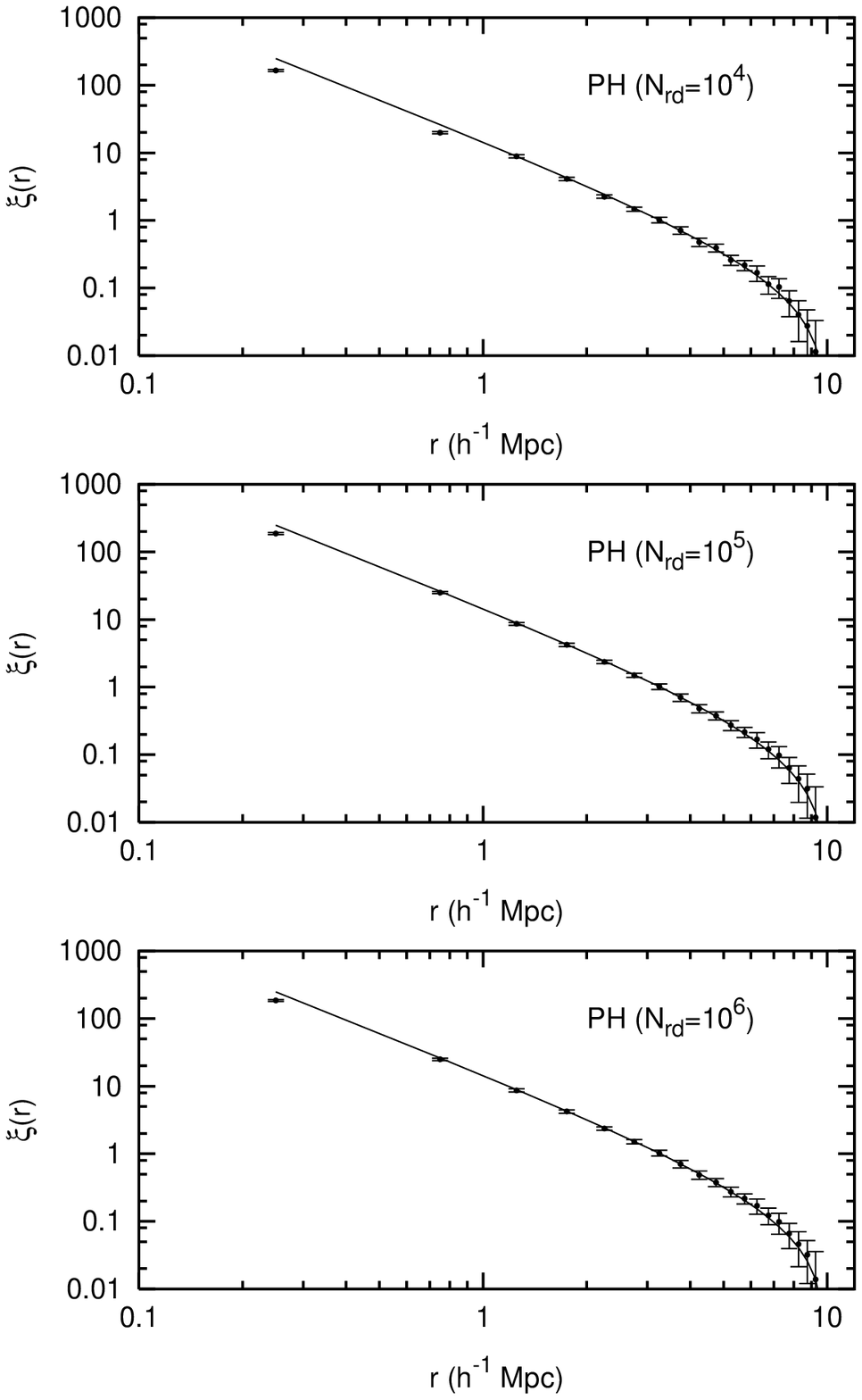}\end{minipage}
\epsfxsize=6.5cm
\begin{minipage}{\epsfxsize}\epsffile{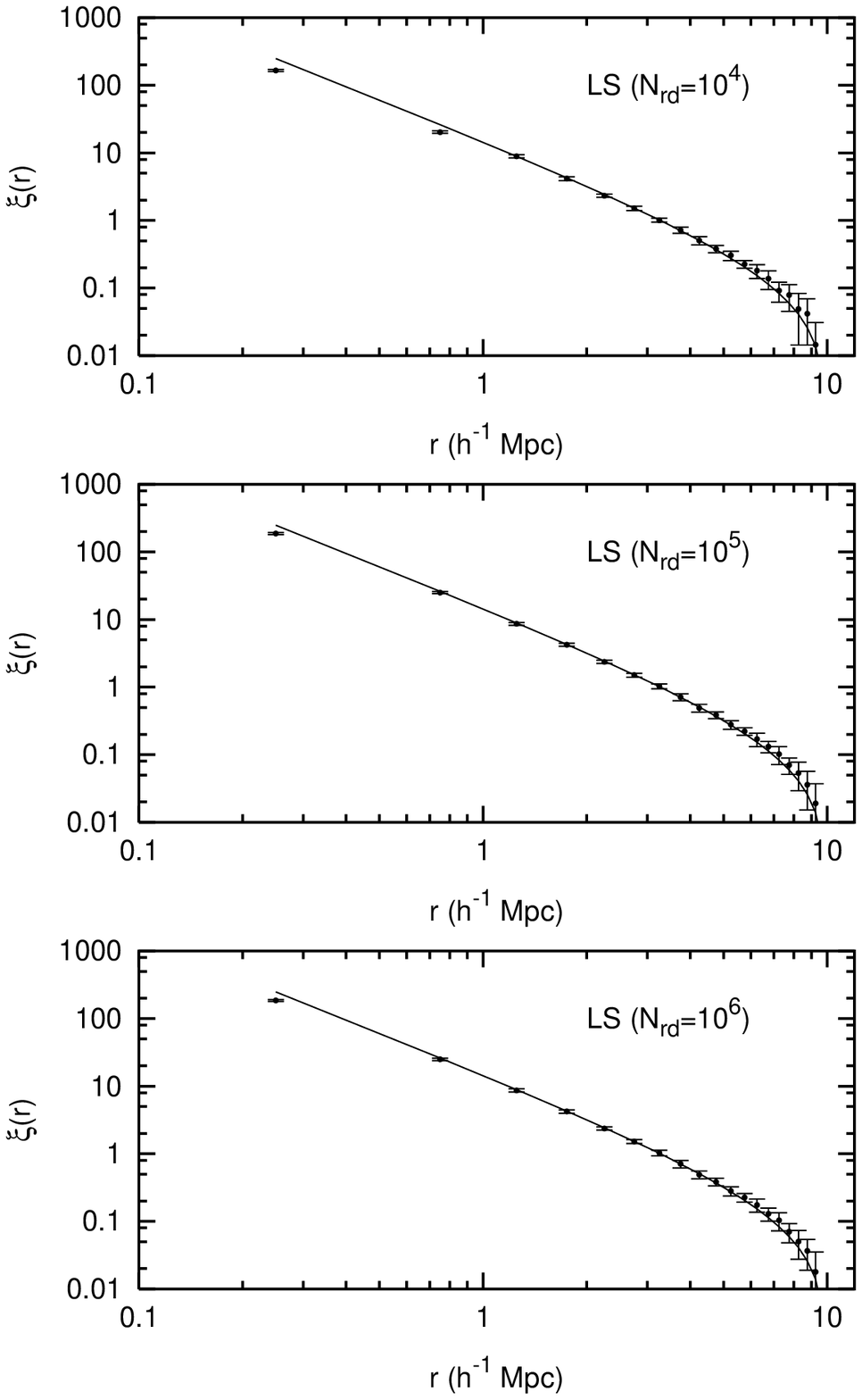}\end{minipage}
\end{center}
\caption{2--point correlation function and standard deviation when
  calculated on 10 realizations of a Cox process for the estimator and
  number of random points indicated in each panel. The continuous line
  corresponds to the analytical $\xi_{\rm {Cox}}$ (Eq. 17).}
\label{error_xcox}
\end{figure*}

First we have performed a couple of tests on 10 Cox processes of the
kind described in Section 4.1, consisting in calculating for them $\xi$ and the
ensemble error with the four estimators introduced in Section 2
depending on $N_{\rm {rd}}$.
We see in Fig.~\ref{error_xcox} what happens when
we increase
the number of random points: $10^4, 10^5, 10^6$.
Our aim is to check if the value of
$N_{\rm {rd}}$  is the source of the differences among them.
In Fig.~\ref{error_xcox} the results of $\xi$ for very small distances
have been suppressed since the
use of Poisson samples introduces shot noise in the estimators because
the local fluctuations become important. One sees that increasing the number of
random points helps reducing the variances, but of course for using
a very large number of
random points, one has to resort to efficient searching algorithms
like those based on the multidimensional binary tree (Mart\'{\i}nez
et al. 1990) to count the number of pairs $RR(r)$ and $DR(r)$.
Alternatively, one has to use analytical expressions for the evaluation
of these quantities (see the appendix in Hamilton (1993)).

Except for the first bin in DP and HAM, the results are practically
the same using $N_{\rm {rd}}=10^5$ than using $N_{\rm {rd}}=10^6$;
that means that, for this process and choice of parameters, 
$N_{\rm {rd}}=10^5$ is ``big enough''. Let us notice that in this
case the difference between PH and LS is very small, tending to 0
as $N_{\rm {rd}}$ increases, since then $(DR(r)/RR(r))
\times(N_{\rm rd}/N)$ tends to 1 (see Eq. 14).

As we can see, 
DP and HAM
estimators have a larger scatter
for the correlation function at short distances than do PH and LS.
This is due to the fact that the shot noise acts to create spurious
clustering in the random samples at small distances,
influencing
the computational number of pairs $DR(r)$ and $RR(r)$
and through those the estimators HAM and DP.
The bigger problem is $DR(r)$ which does not enter in the estimator PH.
If one wishes to use $DR(r)$ as a background number of pairs to normalize
the quantity $DD(r)$, one has to use a large enough random sample in order
to make the fluctuations negligible. But, how large? The intensity (number 
density) of the random sample should be at least that of
the local intensity of the real catalog in the clustered
regions. For example, for the segment Cox processes used here, we 
deduce {\it a priori} the number of random points needed to 
estimate reliably $\xi(r)$ at small separations. From the 
expression of the
correlation function given in Eq.~\ref{skm}, we  know that for this
kind of process the average density at a distance 0.3 $h^{-1}$ Mpc of a given
point is 172.5
times the mean number density, $6 \times 10^{-3}$;
therefore if we want to map
these distances with the random sample, we need at least
$\sim 10^6$ random points
in order that the intensity of the random catalog equals
the previous value of the local
density. At this point it is interesting to remark that at the
smallest interpoint separations, the effects of the finite boundaries
on the estimates of $\xi(r)$ are less important than at large scales;
however it is more difficult to
cope with them with this kind of estimators, because one needs to use
a huge amount of random points or other sophisticated solutions to
get reliable results.

Another practical rule to decide if the random catalogue used
is large enough is to repeat the calculations using different
random seeds -- if the results differ appreciably in the region
of interest, then it is necessary to increase the size of the random
sample (or to choose another estimator).

At intermediate scale all the estimators give the right result with
moderate error bars whereas at large scales the errors increase
for all estimators.  Therefore, the difficulty  to
obtain accurate estimates of $\xi$ at big distances does not seem to be only
due  to the form of a particular estimator or to the number of random points
used but to the {\sl statistic} itself.
Note, however that we have limited our analysis to scales $r \leq l$;
at the end of Section 5.4 we will compare some estimators at longer
distances by means of simulations of the cluster distribution.

\subsection{Estimation of biases}

\begin{figure}
\epsfbox{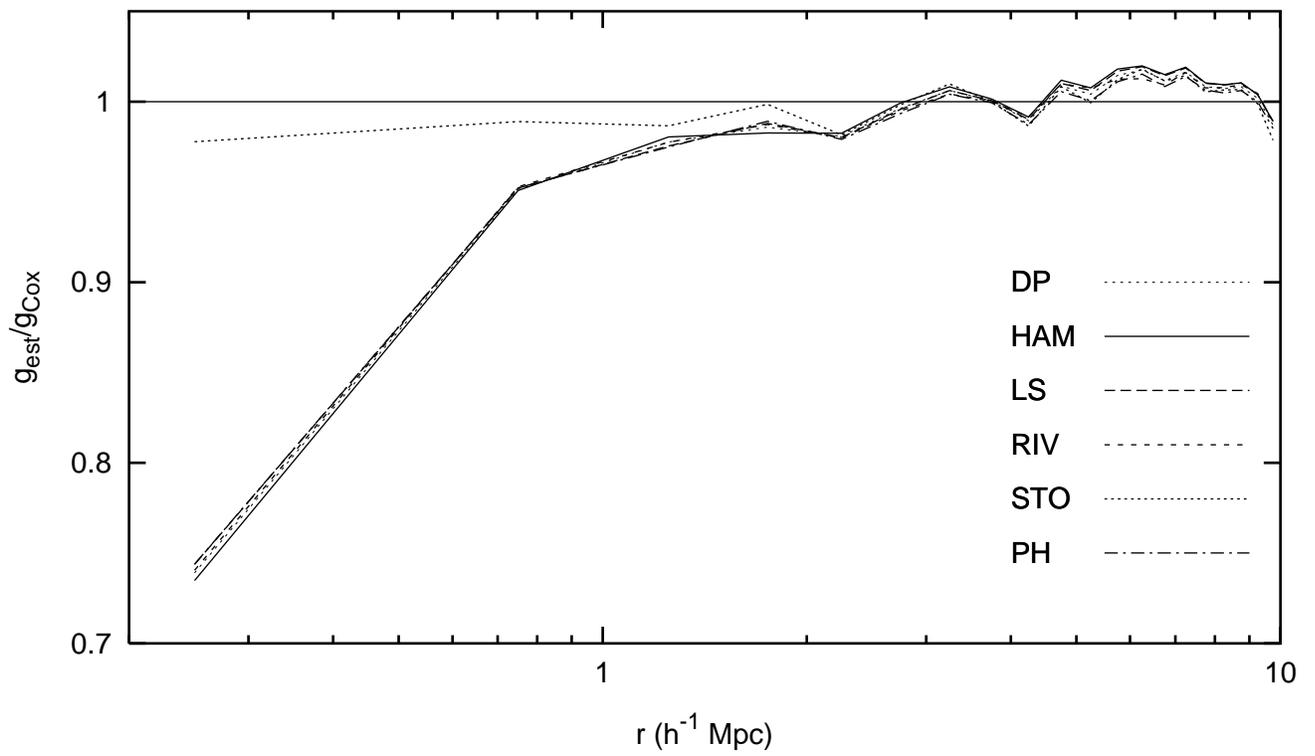}
\caption{Ratio between the mean of $\hat{g}$  and $g_{\rm {Cox}}$, for each
    estimator, calculated on 10 Cox processes using an auxiliary
    random sample containing $N_{\rm {rd}}=10^6$ points.}
\label{bias}
\end{figure}

We shall now consider the results of the previous subsection for
the biggest $N_{\rm{rd}}$ used. Although, as we have seen, increasing
$N_{\rm{rd}}$ reduces the variances, the same effect is not found
for the bias.
We shall proceed to
plot in Fig.~\ref{bias} a measure of the bias in the form of a quotient
between the mean of the 10 estimated
values of $g$ for each estimator using $N_{\rm{rd}}=10^6$, and the
theoretical $g_{\rm {Cox}}$. 
We want also to include for the comparison the STO and RIV estimators.
We shall estimate the volumes entering their definition by means of 
analytical expressions which are available for this simple geometry.

At distances $r \ge 2$\h Mpc
the biases of all the methods are of the same order and the
results for $g(r)$ are quite reliable when compared with the
expected theoretical values given in Eq.~\ref{skm}.
At short distances
the estimator STO performs very well providing the smallest bias.
This good performance is probably related with the fact 
that the segment Cox process
is at small scales locally anisotropic (points randomly placed on a segment)
and as we have explained the STO estimator
deals well with this kind of process. The other estimators show a
clear bias at small scales, underestimating the true value of the
correlation function. It is
expected that for very large windows and a large number of points in
the point sample all estimators are of a similar quality (Hermit et
al. 1996).

\subsection{Variance at large scales}

\begin{figure}
\epsfbox{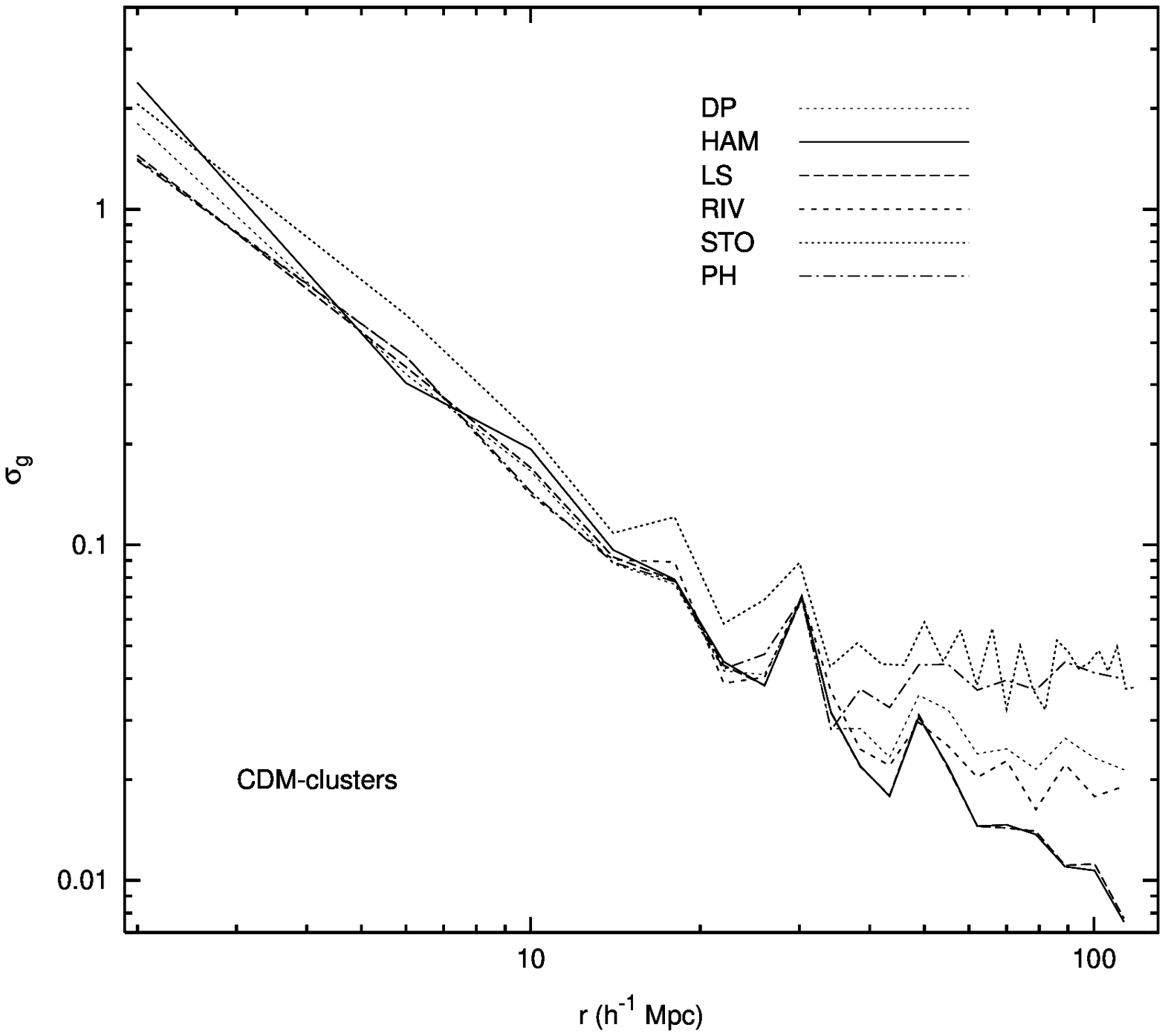}
\caption{Standard deviation of the six estimators when applied to
a set of 10 CDM cluster simulations.}
 \label{sigmacdm}
\end{figure}

The variance for an estimator on a Cox
process could be different from that of the same estimator applied
to galaxy catalogues or cosmological simulations.
Moreover,
the kind of Cox process used here has a limitation due to the finite
length of the segment employed to generate the point distribution, namely
that $\xi$ vanishes for a distance greater than that length.
In order
to see what happens in the absence of such limitation we have taken 10
CDM cluster simulations produced by Croft \& Efstathiou (1994) and
calculated $g$ on
them using the six estimators. The results of their standard
deviation show in Fig.~\ref{sigmacdm} that, at large scale, HAM and LS
have a smaller variance than the others, which could not have been
appreciated in the Cox processes where we should not go farther than 10\h Mpc
in distance. This result supports Hamilton's claim that the estimator
proposed by him (Hamilton 1993) is more reliable on large scales, where the
correlation function is small. Its use provides interesting clues
on the transition to homogeneity of the galaxy distribution at large
scales (Mart\'{\i}nez, 1999).
Other tests
have been performed on simulations for which $g(r)=1$ at
large scales. For these simulations, Hamilton's estimator has
a small systematic bias but a very little estimation variance. Combining
both quantities in the square deviation of the true value, HAM
shows a large degree of precision at large scales.
The reason for that lies in the fact that the term $DR(r)$ in Eq.~\ref{eham}
is related to an improvement of the estimator of the intensity
(Stoyan \& Stoyan 1998).

\section{Estimation of errors using Cox processes}

After having performed the previous tests, we are now ready to use Cox
processes
for estimating errors. We shall do it on the extracted galaxy sample
corresponding to the $\ell=12$ level but the method would be analogous in
the other cases.

As Hamilton (1993) points out (see references therein), five methods
of estimating the variance of $\xi$ are commonly used: Poissonian
error, {\sl idem} enhanced by a certain factor, bootstrap, ensemble
error coming from calculating $\xi$ in subregions of the sample and,
finally, ensemble error coming from artificial samples suffering the
same selection effects than the real sample. The kind of error we are
going to give belongs to the fifth group.

We simulate 10 Cox segment point fields with the 
following values of the parameters
$l =20$ and $\lambda_{\rm{s}}=4\times10^{-5}$. This leads
to a correlation
function
which is comparable with the 2--point correlation
function of the sample of simulated galaxies stopping at the $\ell=12$ level
described in Section 4.2
and which approximately verifies
$\xi(20) =0$ and
$\xi(10)=1$.
Typically these point fields will be generated inside a cube of
80\h Mpc sidelength containing about 800 points.
Using 
similar kind of processes (objects homogeneously distributed in filaments 
and sheets), Buryak \& Doroshkevich (1996) have simulated the galaxy 
distribution.

As can be appreciated in the plots of Fig. 9,
the use of different estimators
causes variability in the slopes of the correlation function.
A least squares fit to a
power--law
for $g(r)\simeq r^{-\gamma}$ in the range [0.5,8] \h Mpc
gives the following results for four of the methods:
$\gamma_{\rm{DP}} = 2.14 \pm 0.06$,
$\gamma_{\rm{HAM}} = 2.27 \pm 0.09$,
$\gamma_{\rm{RIV}} = 2.03 \pm 0.04$,
$\gamma_{\rm{STO}} = 2.03 \pm 0.04$
for a true value $\gamma \simeq 2$
due to the shape of the
power--spectrum (Eq.~\ref{power}).
The fit has been performed
using linear bins and the value of $\hat{g}$ in a particular bin
has been assigned to its centre.
In this case the error accompanying the previous numbers comes from the
weighted least squares fit taking as errors for $g(r)$ the ones
obtained using the Cox processes mentioned in the previous paragraph.

Apart from using these simulations to test the stability of the
methods, we want to stress that this is a way to evaluate the
errors of the correlation function for a given realization,
alternative to the standard bootstrap. Let us stress the idea
of the method, which is 
similar to measuring the dispersion of $\xi$
in ensembles of many independent synthetic catalogues with similar
statistical properties (Fisher et al. 1993):
we use cluster point processes with the same intensity as our sample
and with a known
analytical expression for $\xi(r)$, we build a model having
similar correlation behaviour to that of our galaxy sample, i.\,e.,
a similar $\xi(r)$ in the whole range of scales, and then we are
able to estimate the ensemble error by constructing several
realizations of the point process, applying the estimator of
$\xi$ to all these realizations and measuring the standard
deviation. We believe that this method for the estimation of the
errors is more reliable than the standard bootstrap because of
a serious conceptual weakness the latter suffers from,
namely that the bootstrap suggested in
Ling et al. (1986) produces new point patterns by sampling with replacement;
consequently, in each new point pattern there are multiple points,
i.e., quite heavy clusters. In cluster point
processes the degree of clustering will increase. This leads to incorrect,
probably too great, error predictions. Fisher et al. (1994) show how
bootstrap errors are in general an overestimate of the true errors.

\section{Conclusions}

In this paper we have performed a comparison, by using Cox processes,
of most of the
existing 2--point correlation function estimators.

We would like to point out that a clear distinction has to be
made among the statistical quantity $\xi(r)$, the estimator used
to evaluate it on a particular galaxy catalog, $\hat{\xi}(r)$,
and the particular algorithm of computation of the quantities entering
into the estimator.
It is important to note that what we have compared here is
the performance of different estimators, each implemented in its
simplest way, following the definitions given in Section 2. These kinds
of implementations are the ones commonly used in Cosmology.
In particular,
the estimators depending on the background pair counts $RR(r)$ and
$DR(r)$ need a large amount of random points $N_{\rm{rd}}\sim 10^6$ if
one is trying to accurately measure the correlation function at the
smallest separations, although good enough results can be obtained at
medium and large scales with $N_{\rm{rd}} \sim 20000$.
Note that these figures are appropriate for samples with this density 
but that, for samples with other characteristics, one should previously 
perform tests in order to decide which is a good value for $N_{\rm{rd}}$.
Cox processes are a good benchmark for such tests.
The results show that at large distances all estimators present
similar values and big errors
with HAM and LS clearly being better than the others,
at intermediate
distances values and errors are similar and perfectly acceptable, and at
short distances
the errors for STO are clearly the smallest.
Note, however, that the variance of the former gets smaller
by increasing the number of random points or using alternative
ways for accurately estimating the number of background pairs.
Another advantage of RIV and STO is that they compute something
as easy to accurately estimate as volumes
(in the Monte Carlo implementation the dependence on $N_{\rm {rd}}$ is
softer than for the others because the
random points are being used only for the evaluation of
volumes and not for computation of pairs), whereas in order to increase
accuracy in the others one should make use of a ``big enough'' random
sample and the decision about how big that should be, 
in the absence of previous numerical tests, 
is somewhat arbitrary.
Unfortunately one factor of arbitrariness is always present, namely
the length of the bin in distance (or the coefficient $c$ in the choice
of bandwidth for STO estimator).

The main conclusion we have drawn from our analysis is that there
exists no optimal estimator but that each one has advantages and weak
points and, depending on the
nuances of the problem we want to analyze,
one or another will be preferable. In the case of complete samples
limited in volume, RIV is not very sensitive to the
number of random points used to evaluate the volumes but presents a
bias at small distances; HAM has small variance at long distances but
larger at small distances and in this range
is highly sensitive to $N_{\rm{rd}}$ and it is biased; DP has a big variance
and presents a bias at short scales;
PH depends also on $N_{\rm{rd}}$ but less than HAM and DP and
also shows a bias at small scales; STO is never the worst
in any of the tests and can be applied also to anisotropic processes;
and LS behaves in many aspects similarly to PH but with a smaller variance
at large scale.
For samples with non-uniform density these conclusions may vary,
and in particular HAM is preferable at large scales.

Two further points---secondary with regard to the comparison of estimators
but also interesting and potentially
useful for researchers on this field---have been treated:
for testing the estimators we have introduced a new phenomenological
method to extract galaxy samples from cosmological simulations
based on
the multidimensional binary trees; and, for such samples, we have
estimated errors in the determination of the 2--point correlation
function by using realizations of a Cox process with the same number
density as the simulated sample.

\acknowledgements
This work was partially supported by
the Spanish DGES project n. PB96-0797.
D. Stoyan was partially supported by a grant of the
Deutsche Forschungsgemeinschaft.
E. Saar acknowledges a fellowship of the Conselleria de Cultura, Educaci\'o
i Ci\`encia de la Generalitat Valenciana.
We are grateful to R. Croft, S. Hermit, M. Graham
and J. Loveday for kindly permitting us to use part of their
data and results.
Advice from M. Stein and R. Moyeed is also acknowledged.
We thank the referee, Douglas Tucker, as well as Andrew Hamilton and Martin 
Kerscher for a careful reading of the manuscript and for their valuable 
comments and suggestions.

\newpage

\begin{thebibliography}{99}
\bibitem[Baddeley et al. 1993]{rana} Baddeley A.J., Moyeed R.A.,
Howard C.V., Boyde A., 1993, Appl. Statist., 42, 641
\bibitem [Blanchard \& Alimi 1988]{blan88}  Blanchard, A., Alimi,
J.-M., 1988, A\&A, 203, L1
\bibitem[]{bon94} Bonometto A.A., Iovino A., Guzzo L., Giovanelli R.,
Haynes M., 1994, ApJ, 419, 451
\bibitem[]{bur95} Buryak O., Doroshkevich A., 1996, A\&A, 306, 1
\bibitem[Croft \& Efstathiou 1994]{cro94} Croft R.A.C., Efstathiou G., 1994,
MNRAS, 268, L23
\bibitem[]{dal94} Dalton, G.B., Croft R.A.C., Efstathiou G.,
Sutherland, W.J., Maddox, S.J., Davis, M., 1994, MNRAS, 271, L47
\bibitem[]{dp83} Davis M., Peebles P.J.E., 1983, ApJ, 267, 465
\bibitem[]{dav88} Davis M., Meiskin A., Strauss M.A., da Costa L.N., Yahil A.,
1988, ApJ, 333, L9
\bibitem[]{del86} de Lapparent V., Geller M.J., Huchra J.P., 1986, ApJ,
302, L1
\bibitem[]{del88} de Lapparent V., Geller M.J., Huchra J.P., 1988, ApJ,
332, 44
\bibitem[Doguwa \& Upton 1986]{dog86} Doguwa, S.I., Upton, G.J.G., 1986,
  Biom. J., 31, 563
\bibitem[]{fis93} Fisher K.B., Davis M., Strauss M.A., Yahil A., Huchra J.P.,
1993, ApJ, 402, 42
\bibitem[]{fis94} Fisher K.B., Davis M., Strauss M.A., Yahil A., Huchra J.P.,
1994, MNRAS, 266, 50
\bibitem[Hamilton 1988]{ham88} Hamilton A.J.S. 1988, ApJ, 331, L59
\bibitem[Hamilton 1993]{ham92} Hamilton A.J.S. 1993, ApJ, 417, 19
\bibitem[]{her96} Hermit S., Santiago B.X., Lahav O., Strauss M.A.,
Davis M., Dressler A., Huchra J.P., 1996, MNRAS, 283, 709
\bibitem[Kerscher et al]{aiha} Kerscher, M., Pons--Border\'{\i}a, M.J.,
Schmalzing, J., Trasarti--Battistoni, R., Buchert, T.,
Mart\'{\i}nez, V.J. \& Valdarnini, R. 1999, ApJ, 513, 543
\bibitem[Kerscher 1999]{ker98} Kerscher, M., 1999, A\&A, 343, 333
\bibitem[]{lan93} Landy S.D., Szalay, A.S., 1993, ApJ, 412, 64
\bibitem[]{lov95} Loveday J., Maddox S.J., Efstathiou G.,
Peterson B.A., 1995, ApJ, 442, 457
\bibitem[]{mad90} Maddox S.J., Efstathiou G., Sutherland W.J., Loveday J.,
1990, MNRAS, 242, 43{\small P}
\bibitem[Mart\'{\i}nez et al. 1990]{mar90}  Mart\'{\i}nez, V.J.,
Jones, B.J.T., Dom\'{\i}nguez-Tenreiro, R., \& van de Weygaert, R. 1990,
ApJ 357, 50
\bibitem[]{mar93} Mart\'{\i}nez V.J., Portilla M., Jones B.J.T.,
Paredes S., 1993, A\&A, 280, 5
\bibitem[]{mar94} Mart\'{\i}nez V.J., Coles P., 1994, ApJ, 437, 550
\bibitem[]{mar99} Mart\'{\i}nez V.J., 1999, Science 284, 445
\bibitem[]{mau92} Maurogordato S., Schaeffer R., da Costa L.N., 1992,
ApJ, 390, 17
\bibitem[]{moo94} Moore B., Frenk C.S., Efstathiou G., Saunders W., 1994,
MNRAS,
269, 742
\bibitem[]{lin86} Ling E.N., Frenk C.S., Barrow J.D., 1986, MNRAS, 223, 21
\bibitem[]{par94} Park C., Vogeley M.S., Geller M.J., Huchra J.P., 1994,
ApJ, 431, 569
\bibitem[]{pea92} Peacock, J.A., 1992, in Mart\'{\i}nez, V.J.,
Portilla, M., \& S\'aez, D. Ed., New Insights into the Universe,
Springer--Verlag, Berlin
\bibitem[Peebles 1980]{lss} Peebles, P.J.E., 1980, The Large Scale Structure
of the Universe, Princeton University Press, Princeton
\bibitem[Peebles \& Hauser 1974]{pee74} Peebles, P.J.E.,
  Hauser, M.G. 1974, ApJ Suppl., 28, 19
\bibitem[Ripley 1981]{rip81}  Ripley, B.D. 1981, Spatial Statistics,
John Wiley \& Sons, New York
\bibitem[]{riv86} Rivolo A.R., 1986, ApJ, 301, 70
\bibitem[Santiago et al. 1995]{ors} Santiago, B.X.,
  Strauss, M.A., Lahav, O., Davis, M., Dressler, A.,
  Huchra, J.P., 1995, ApJ, 446, 457
\bibitem[Santiago et al. 1996]{ors2} Santiago, B.X.,
  Strauss, M.A., Lahav, O., Davis, M., Dressler, A.,
  Huchra, J.P., 1996, ApJ, 461, 38
\bibitem[]{sau92} Saunders W., Rowan-Robinson M.,
  Lawrence A., 1992, MNRAS, 258, 134
\bibitem[Stoyan \& Stoyan 1994]{sto94} Stoyan, D., Stoyan, H., 1994,
Fractals, Random Shapes and Point Fields, J. Wiley \& Sons, Chichester
\bibitem[Stoyan \& Stoyan 1996]{sto96} Stoyan, D., Stoyan, H., 1996,
Biom. J., 38, 259
\bibitem[Stoyan et al. 1995]{sto95} Stoyan, D., Kendall, W.S., Mecke, J.,
1995, Stochastic Geometry and its Applications, J. Wiley \& Sons, Chichester
\bibitem[Stoyan \& Stoyan 1998]{sto98} Stoyan, D., \& Stoyan, H. 1998,
preprint 98--3 Technische Universit\"at Bergakademie Freiberg
\bibitem[Szapudi \& Szalay 1997]{sza97} Szapudi, I., \& Szalay, A.S.,
1998, ApJ, 494, L41
\bibitem[Suisalu et al. 1999]{saar} Suisalu, I., Saar, E., Jones, B.J.T.,
1999, in preparation
\bibitem[Tucker et al 1997]{tuc97} Tucker et al., 1997, MNRAS, 285, L5
\bibitem[van de Weygaert 1988]{rien} van de Weygaert, R. 1988, Master Thesis,
Rijksuniversiteit Leiden
\bibitem[van de Weygaert 1991]{van88} van de Weygaert, R. 1991, MNRAS, 249, 159
\end{thebibliography}
\end{document}